# Temporal Fresnel diffraction induced by phase jumps in linear and nonlinear optical fibres

Anastasiia Sheveleva and Christophe Finot *

*Laboratoire Interdisciplinaire Carnot de Bourgogne, UMR 6303 CNRS-Université Bourgogne Franche-Comté, Dijon, France*

* christophe.finot@u-bourgogne.fr

# Temporal Fresnel diffraction induced by phase jumps in linear and nonlinear optical fibres

Abstract  We analytically and numerically study the temporal intensity pattern emerging from the linear or nonlinear evolutions of a single or double phase jump in an optical fiber. The results are interpreted in terms of interferences of the well-known diffractive patterns of a straight edge, strip and slit and a complete analytical framework is provided in terms of Fresnel integrals for the case of purely dispersive evolution. When Kerr nonlinearity affects the propagation, various coherent nonlinear structures emerge according to the regime of dispersion.

Keywords: space/analogy ; dispersion ; nonlinear fiber optics ; Fresnel diffraction

## Introduction

Diffraction is among the key effects of wave physics with applications in a broad range of technological domains spanning from imaging to spectroscopy, material sciences, mobile communications, test measurements or sensors. Propagation of a plane wave through a sharp edge is the unavoidable example taught in any physical optics courses to introduce the concepts and the theoretical tools available to handle Fresnel's diffraction [1, 2]. But diffraction is not restricted to opaque screens that partially obstruct light, it is also involved when light is transmitted through a phase plate, i.e. a transparent medium imprinting a localized phase jump, leading to strong oscillations of the diffracted field [3, 4].

The well-known free-space evolution of a beam can find analogs in the temporal domain. Indeed, the parabolic spectral phase induced by the dispersion of an ultrashort optical pulse is equivalent to the paraxial diffraction affecting the spatial propagation of a light beam [5-7]. This space/time duality has already been extremely fruitful and has stimulated numerous new concepts or interpretation in ultrafast optics such as temporal or spectral lenses [8, 9], Fresnel lens [10], super resolution imaging [11], dispersion gratings [12] or two-wave temporal interferometers [13], to cite a few. Other studies have established links between the near-field propagation observed in diffraction and advanced applications for high repetition-rate sources when initial periodic phase modulation is converted into intensity pattern [14-18] in a process that can be linked to the Talbot array illuminators in the spatial domain [19]. However, despite the fact that coherent communications now heavily rely on the use of phase modulation, no explicit study of the space / time duality for a single phase step has been clearly reported so far.

This is the scope of the present paper to fill this gap by providing a series of analytical and numerical results for a single phase jump evolving in a single mode optical fiber. We then extend the discussion to the case of a double phase jump and highlight some significant differences with respect to the pattern resulting from the dispersion of a temporal hole of light. In both cases, we also investigate the consequences of optical Kerr nonlinearity according to the regime of dispersion and demonstrate the emergence of coherent structures. Finally, the impact of the finite bandwidth of temporal modulation is discussed.

**Study of a single and ideal temporal phase jump**

*Situation under study and analytical treatment of the linear propagation*

Before discussing our experiments, let us first recall the basis of the analogy between the spatial evolution of light affected by diffraction and the temporal changes experienced by light when dispersion is involved. We consider the simple case where a monochromatic plane wave with wavelength $\lambda$ and an amplitude $a_0$ illuminates a phase pattern. It can be a light beam transmitted through a transparent plate with an abrupt change in thickness or refractive index. For a 1D transverse problem, the longitudinal evolution of light $a(x,z)$ in the scalar approximation is ruled by the following differential equation :

$$i\frac{\partial a}{\partial z} = -\frac{1}{2k_0}\frac{\partial^2 a}{\partial x^2}, \tag{1}$$

with $x$ and $z$ being the transverse and longitudinal coordinates respectively and $k_0 = 2\pi/\lambda$ the wavenumber. The goal of the present paper is to study the temporal equivalent of this configuration. We therefore consider an initial continuous wave $a(t, z=0)$ where light has been modulated by an abrupt phase offset $\Delta\varphi$ at $t = 0$ that follows :

$$\begin{cases} a(t,0) = 1 & \text{for } t \leq 0 \\ a(t,0) = \exp(i\,\Delta\varphi) & \text{for } t > 0 \end{cases} \tag{2}$$

with $t$ being the temporal coordinate. This shaped temporal waveform then propagates in a dispersive single mode waveguide, typically an optical fiber, that ensures that its spatial transverse profile is unaffected upon propagation. The temporal profile of the light in the approximation of the slowly varying envelope evolves according to:

$$i\frac{\partial a}{\partial z} = \frac{1}{2}\beta_2\frac{\partial^2 a}{\partial t^2}, \tag{3}$$

with $\beta_2$ the second order dispersion coefficient. We have here considered a second order anomalous dispersion $\beta_2 = -20$ ps$^2$/km typical of the SMF-28 fiber used for optical telecommunications [20]. As this second-order dispersion coefficient is much higher than the third-order dispersion coefficient, it is possible, as a first approximation, to fully neglect the impact of higher order dispersive terms. The space-time duality readily appears in the mathematical structure of equations (1) and (3) where the transverse space coordinate and time are exchanged, the waveforms fulfilling the same normalized differential equation [5-7]. Consequently, both diffraction and dispersion imply the development of a quadratic spectral phase and lead formally to similar consequences. In order to better understand the evolution of temporally phase-sculpted waveform subject to dispersion, given the superposition property, it may be useful to rewrite the initial condition of our problem given by Eq. (2) as a linear combination of two patterns illustrated in Fig. 1(a) : $a(t,0) = \psi_1(t) + \psi_2(t)$, with $\psi_1(t) = H(-t)$ (green line) and $\psi_2(t) = H(t) \exp(i\,\Delta\varphi)$ (purple line), $H$ being the Heaviside step distribution that represents in the 1D spatial domain the equivalent of a straight edge. Therefore, the pattern that emerges upon linear propagation can be viewed as the result of the interference between the typical waveforms linked to the diffraction of two out-of-phase straight edges $\psi$ of opposite orientations. Let us remind that the diffraction pattern of an ideal semi-infinite screen $\psi$ is among the first examples that are taught to students when introducing diffraction [1, 2, 21]. For this case, Fraunhofer approximations do not hold but the edge problem can be solved analytically by involving Fresnel's integrals $C_f$ and $S_f$ and the graphical Cornu's spiral plot:

$$\psi(x,z) = \frac{1}{2}\left\{ C_f\left(\sqrt{\frac{2}{\lambda z}}x\right) + 0.5 + i\left[S_f\left(\sqrt{\frac{2}{\lambda z}}x\right) + 0.5\right]\right\} \quad (4)$$

Expressed in the context of temporal evolution, this leads to:

$$\psi(t,z) = \frac{1}{2}\left\{ C_f\left(\frac{\mathrm{sgn}(\beta_2)\,t}{\sqrt{\pi|\beta_2|z}}\right) + 0.5 + i\left[S_f\left(\frac{\mathrm{sgn}(\beta_2)\,t}{\sqrt{\pi|\beta_2|z}}\right) + 0.5\right]\right\}. \quad (5)$$

The temporal intensity profile of one semi-infinite edge is therefore characterized by strong oscillations of the plateau as can be seen in Fig. 1(b) with extrema located at $\pm t_{exp,m}$ ($m \geq 0$) given by :

$$t_{ext,m} = \sqrt{2\pi\left(\frac{3}{4}+m\right)|\beta_2|\,z} \quad (6)$$

The most pronounced ripple is obtained at $t_{max} = t_{ext,0}$ and has a maximum equal to 1.37 which is independent of the propagation distance. It is followed by a local minimum at $t_{min} = t_{ext,1}$.

The temporal phase difference $\Delta_T\varphi$ between $\psi_1$ and $\psi_2$ can be expressed as $\Delta_T\varphi(t,z) = \arg(\psi_2(t,z)) - \arg(\psi_1(t,z)) = \arg(\psi(t,z)) - \arg(\psi(-t,z)) + \Delta\varphi = \delta\varphi + \Delta\varphi$ with $\delta\varphi = \arg(\psi(t,z)) - \arg(\psi(-t,z))$ plotted with black line in panel (b2) and that can be analytically derived as :

$$\tan(\delta\varphi(t,z)) = \text{sgn}(\beta_2) \frac{C_f(u) - S_f(u)}{C_f^2(u) + S_f^2(u) - 1/2} \tag{7}$$

with the normalized coordinate $u$ being $u(t) = \text{sgn}(\beta_2) t / \sqrt{\pi |\beta_2| z}$. One can note that $\delta\varphi$ is null at $t = 0$ and is close to an even multiple of $\pi$ around $t_{max}$. This formula can be closely adjusted around $t = 0$ by the following linear fit (black dashed line) :

$$\delta\varphi(t,z) = -\text{sgn}(\beta_2) \sqrt{\frac{\pi}{2|\beta_2| z}} \, t \tag{8}$$

The intensity profile obtained after propagation is given by the following expression that can be interpreted exploiting Cornu's spiral [4, 22] :

$$|a(t,z)|^2 = \cos^2(\Delta\varphi/2) + 2\left[C_f^2(u) + S_f^2(u)\right]\sin^2(\varphi/2) \\ + \left[C_f(u) - S_f(u)\right]\sin(\Delta\varphi) \tag{9}$$

One particularly interesting case is when $\Delta\varphi$ equals $= \pi$ (see Fig. 2, panel (a) and red curve in panel (b)). In this case, $\Delta_T\varphi(t)$ equals $\pi$ around the central position so that the interference between $\psi_2(t)$ and $\psi_1(t)$ is destructive around $t=0$. As $\psi_2(0)$ and $\psi_1(0)$ have identical intensities, the destructive interference is complete and the intensity drops to zero. Therefore, the intensity in the central part is lowered compared to what would have been expected from the incoherent sum of the intensities of the two diffraction pattern $|\psi_1|^2 + |\psi_2|^2$ (Fig. (2), black line). On the contrary, around $t_{max}$, $\delta\varphi$ becomes an odd multiple of $\pi$, so that the interference becomes constructive and the peak intensity is enhanced by 31%. As a consequence, the prediction of Eq. (6) initially derived for an abrupt intensity edge also applies for a step phase edge as can be checked in panel 2(a) (see dashed line). Note that for $\Delta\varphi = \pi$, the expression of the resulting temporal intensity pattern becomes analytically extremely simple [23]:

$$|a_\pi(t,z)|^2 = 2\left[ C_f^2(u) + S_f^2(u) \right], \quad (10)$$

making it straightforwardly interpreted by the use of the Cornu's spiral. The temporal duration $\Delta T$ of the central dip [23] then evolves as:

$$\Delta T \simeq \sqrt{2\pi |\beta_2| z}. \quad (11)$$

with $|a_\pi(t = \pm \Delta T / 2, z)|^2 \simeq 1$ and $\delta\varphi \simeq \pm \pi/2$ so that $|\psi_1|^2 + |\psi_2|^2$ do not interfere (the red and black curves on panel (b) of Fig. 2 have the same value). This symmetric pattern characterized by a central hole surrounded by two marked oscillating edges that move in opposite directions should not be confused with the pattern resulting from a binary intensity modulation of the initial condition as discussed in [24]. Indeed, in this last configuration, a central spot was progressively growing at the centre of the light hole due to constructive interference.

One advantage of temporal optics compared to spatial optics (that requires fine tuning of precision optics [25]) is that it is quite straightforward to adjust optoelectronic devices to modify $\Delta\varphi$. As an example, we have considered $\Delta\varphi = \pi/2$ in panel (b) of

Fig. 2 (blue line). Several features that can be explained once again through the interference process and $\delta\varphi$. First, the intensity pattern is not symmetric anymore with respect to $t = 0$, as can be clearly visible from the value and position of the maxima. Indeed, when $\Delta\varphi = \pi/2$, the maxima are obtained for $\delta\varphi \simeq -\pi/2$ or $3\pi/2$, i.e., for times that are $-18.6$ or $41.2$ ps after 10 km of propagation (see Fig. 1(b2)). The amplitude of the ripple is also affected and the temporal fringes that develop on each side of the dip are not identical: for $\Delta\varphi = \pi/2$, the peak at $t = -35.4$ ps is lowered as the interference process involves a tail of $|\psi_1|^2$ with a reduced intensity. On the contrary, the bump at $t = 24.1$ ps, is increased as the tail linked to $|\psi_2|^2$ is more powerful and therefore stimulates a constructive interference with a higher efficiency. We also note that the intensity does not drop to zero in the central part. The destructive interference condition is obtained for $\Delta\varphi = \pi/2$ (i.e. at $t = \Delta T/2$) so that $\psi_2(t)$ and $\psi_1(t)$ have significantly different values that lowers the efficiency of the destructive process. Those trends are fully confirmed by the evolution of the pulse pattern recorded after 10 km of propagation according to $\Delta\varphi$ and summarized in panel (c) of Fig. 2. We observed the change in the visibility of the central fringe [25] as well as the continuous shift of the maxima. In order to qualitatively predict the temporal location $t_{min}$ of this dip, we can take advantage of the linear approximation of $\delta\varphi$ [Eq. (8)] to propose the following empirical prediction:

$$t_{min} \simeq -\frac{\text{sgn}(\beta_2)}{2}\sqrt{\frac{2|\beta_2|z}{\pi}}(\pi - \Delta\varphi). \quad (12)$$

Panel (c) of Fig. 2 confirm the reasonable agreement of Eq. (12) which is close to linear shift of $t_{min}$ according to the initial amplitude of the phase offset. Note that if the regime of dispersion is normal instead of anomalous, the pattern will be flipped in the temporal domain (see Fig. 2(b), dashed blue curve). We can note that the amplitude of only the first central bumps is affected: for larger time, the intensity of the tails of $\psi_1$ or $\psi_2$ becomes negligible so that the influence of the interference process on the ripple becomes much more negligible.

*Nonlinear propagation*

We are now interested in the propagation occurring when nonlinear effects become significant. Indeed, contrary to the usual diffraction in free space, propagation in a waveguide can also involve nonlinear effects. In this context, optical fibers represent an ideal testbed: thanks to a very low level of losses, the Kerr nonlinearity of silica may be accumulated over several kilometers. The temporal evolution of a waveform in an optical fiber resulting from the interaction between nonlinearity and dispersion can be taken into account through an additional term accounting for self-phase modulation in Eq.(3), leading to the well-known nonlinear Schrödinger equation (NLSE) [26]:

$$i\frac{\partial a}{\partial z} = \frac{1}{2}\beta_2 \frac{\partial^2 a}{\partial t^2} - \gamma |a|^2 a, \qquad (13)$$

with $\gamma$ is the Kerr coefficient of the fiber, typically $\gamma = 1.1$ /W/km for the SMF-28 fiber. As a first approximation, we have neglected the impact of the optical losses that are reduced in the telecommunication spectral window of telecommunication fiber (around 0.2 dB/km) and can be ideally compensated used Raman distributed amplification [27]. We solve the scalar NLSE using numerical simulations based on the well-established split-step Fourier method [26].

The longitudinal evolution of the intensity profile for an initial power of 290 mW is plotted in Fig. 3(a1) for an initial phase offset of π, with details of the intensity profiles recorded after 10 km of propagation in SMF-28 fiber given in panel (b1). When compared to the linear propagation (see Fig. 2(a) and dashed line of Fig 3(b1)), the impact of the nonlinearity is readily visible. Whereas the position of the maximum is still close to $t_{max}$ predicted by Eq. (6), the central gap broadens and the ripples are dramatically compressed. The most compressed structures appear first on the edges of the central gap where the initial high-intensity fluctuations trigger the nonlinear effects. These coherent structures become narrower and narrower and lie on a zero background. They also have a peak intensity that is strengthened, with a maximum value of 1.8 in the linear regime of propagation that now exceeds 4times the input average value in the nonlinear propagation. Such a behavior qualitatively recalls the experimental observations made in the spatial domain for the nonlinear Fresnel diffraction [28]. This dynamic is ascribed to the focusing nature of the nonlinearity in the anomalous regime of propagation which has been recently the subject of many investigations exploring the nature of the solitonic or breathing structures that may develop upon propagation. Those works have emphasized the universal impact of modulation instability and may have proposed mathematical methods [29-32] such as the nonlinear Fourier transform to identify and interpret the fine details of the coherent nonlinear structures. However, the configurations that have been treated have often been restricted to the semi-classical limit of the NLSE and they have essentially considered the case of an initial perturbation of the intensity profile that then translates into large fluctuations amplified by modulation instability process [33-37]. In order to get insights on the nature of the first most intense structure, we can stress that the intensity profile is in excellent agreement with the shape of a fundamental bright soliton $\psi_S(t)$ (with here a term of longitudinal phase offset omitted):

$$\psi_S(t) = \sqrt{P_S}\ \text{sech}\left(\frac{t}{T_S}\right), \qquad (14)$$

with $T_S$ and $P_S$ the temporal duration and peak-power of the fundamental soliton linked by $T_S = (\beta_2 / P_S\ \gamma)^{1/2}$. Figure 3(c1) summarizes the evolution of the pattern for a fixed propagation length of 10 km. We can note that the width of the coherent structures that surround the central gap is also affected by the input power with the degree of compression increasing with the initial power. The pattern seems to asymptotically tend to a modulated soliton train with unequal temporal spacing and with a peak power close to four times the initial average power, as predicted for other initial conditions in [32, 35]. The temporal location of those soliton-like peaks is also affected by the peak power. Whereas the temporal shift induced by nonlinearity is quite limited on the most central solitons (see also dotted white line in panel (a1)), we note that the temporal location of other bright structures emerging from secondary ripples is more power-dependent.

The picture gets very different when normal dispersion is involved (panel (a2), $\beta_2 = 20$ ps$^2$/km). Instead of a central gap broadening with propagation distance according to Eq. (11), we observe a dip that evolves with its temporal duration unaffected, while its minimum goes down to a null intensity. The details of the intensity profiles plotted in Fig. 3(b2) for a propagation distance of 5 and 10 km stress that the width of the central part does not evolve and is good agreement with a black soliton $\psi_{BS}(t)$, which is characterized by a full hole of light and a phase offset of $\pi$ at its center [38], with an analytical expression provided by :

$$\psi_{BS}(t) = \sqrt{P_{BS}}\ \tanh\left(\frac{t}{T_{BS}}\right), \qquad (15)$$

with $T_{BS}$ the temporal duration of the black soliton and $P_{BS}$ the power of the continuous background linked by $T_{BS} = \left(\beta_2 / P_{BS}\, \gamma\right)^{1/2}$. Due to the imperfect initial profiles, this dark soliton is surrounded by radiative waves that progressively move away from the central part. Consequently, it confirms that imprinting an initial temporal phase singularity of π is a possible approach to generate black solitons [39], which contrasts with the others technics that have focused on the advanced shaping of the temporal intensity and phase profile [40, 41] or on the nonlinear interaction of a pair of delayed pulses [42, 43]. Figure 3(c1) points out that the balance between normal dispersion and Kerr nonlinearity lead to black solitons having a central gap that gets narrower and narrower when increasing power, which is fully consistent with the scaling laws of the parameters of Eq. (15).

The depth of the initial phase jump influences the symmetry of the resulting pattern as outlined in panel (c2) of Fig. 3 that summarizes the nonlinear evolution of a π/2 phase step. In the normal regime of dispersion, the initial non vanishing dip that appears upon linear propagation is converted into a grey soliton-like structure that has a reduced contrast. Both the reduced initial phase offset as well as the reduced depth of the main dip induced by dispersion contribute to generate a grey soliton with a reduced greyness [26]. Note that the grey soliton has a non-zero velocity, as also stressed in spatial optics [44]. Regarding the evolution recorded in the anomalous dispersion regime for $\Delta\varphi = \pi/2$, we note that the fluctuations of peak power and widths of the localized ultrashort structures according to the input power becomes pronounced. This breathing behavior is dominated by solitons over finite background such as Akhmediev breathers, Kutnetsov-Ma solutions or superregular structures [45] that may exist both in temporal optics [46] but also in spatial optics [47]. It is worthy to note that the highest peak power is not always achieved for negative times as we could have expected from the asymmetry existing in the linear propagation.

**Study of the double phase jump**

*Situation under study and dispersive propagation*

The diffraction of 2D transparent phase objects such as square or circular samples have been the subjects of past investigations in spatial wave optics, with applications to metrology [4]. We now consider the case where the temporal phase shift imprinted on the continuous wave $\Delta\varphi$ is limited to a duration $T_0$. The spatial analog of this initial condition is a 1D is a transparent stripe of width $x_0$ with a height leading to a phase offset $\Delta\varphi$. This stripe has two abrupt edges and is illuminated in normal incidence. Our ideal temporal object can be analytically described as:

$$\begin{cases} a(t,0) = 1 & \text{for } |t| \geq T_0/2 \\ a(t,0) = \exp(i\,\Delta\varphi) & \text{for } |t| < T_0/2 \end{cases}. \quad (16)$$

Once again, we can take advantage of the superposition principle to facilitate the discussion of the dispersion phenomena and the analytical calculations [25, 48]. Indeed, there are various ways to rewrite Eq. (16). One can see this problem as the temporal coherent addition of two abrupt phase jumps of similar amplitude but with opposite temporal orientations overlapping by the duration $T_0$. Another way to interpret the initial condition is illustrated in Fig. 4(a). It is convenient to rewrite this initial condition as the sum of three elements: $a(t,0) = \psi'_1(t) + \psi'_2(t) + \psi'_3(t)$, with $\psi'_1(t) = H(-t+T_0/2)$, $\psi'_2(t) = H(t-T_0/2)$ and $\psi'_3(t) = \text{rect}(t/T_0)\exp(i\,\Delta\varphi)$, rect being the rectangular function of width 1 (purple curve). The pattern made by $\psi_A(t) = \psi'_1(t) + \psi'_2(t)$ (green curve) corresponds to an opaque light hole of width $T_0$. This temporal analogue of an opaque stripe of constant width [49-51] has been the subject of our recent paper dealing with the observation of the temporal Arago spot in optical fibers [24]. Therefore, our problem resumes to the coherent superposition of the temporal Arago pattern and the temporal pattern induced by an aperture of width $T_0$ with a phase offset of $\Delta\varphi$. The intensity profiles linked to both waves can be analytically derived [48] and are plotted on Fig. 4(b1) for a propagation distance of 10 km. We observe the central intensity bump typical of the Arago spot that will interfere with the temporally broadened pattern induced by the rectangular phase offset. The problem can be solved analytically and the temporal profile can be once again predicted using a combination of Fresnel integrals:

$$
\begin{aligned}
|a(t,z)|^2 = &\ 1 \\
&+ \left[C_f(u_1) - C_f(u_2)\right]\left\{\left[C_f(u_1) - C_f(u_2)\right](1 - \cos\Delta\varphi) - 1 + \cos\Delta\varphi - \sin\Delta\varphi\right\} \\
&+ \left[S_f(u_1) - S_f(u_2)\right]\left\{\left[S_f(u_1) - S_f(u_2)\right](1 - \cos\Delta\varphi) - 1 + \cos\Delta\varphi + \sin\Delta\varphi\right\}
\end{aligned} \quad (17)
$$

with $u_1(t) = \mathrm{sgn}(\beta_2)(t - T_0/2)/\sqrt{\pi|\beta_2|z}$ and $u_2(t) = \mathrm{sgn}(\beta_2)(t + T_0/2)/\sqrt{\pi|\beta_2|z}$.

This can be further simplified in the case of $\Delta\varphi = \pi$ to :

$$
\begin{aligned}
|a_\pi(t,z)|^2 = &\ 1 + 2\left[C_f(u_1) - C_f(u_2)\right]\left[C_f(u_1) - C_f(u_2) - 1\right] \\
&+ 2\left[S_f(u_1) - S_f(u_2)\right]\left[S_f(u_1) - S_f(u_2) - 1\right]
\end{aligned} \quad (18)
$$

The phase difference $\delta\varphi' = \arg(\psi_3(t)) - \arg(\psi_A(t)) - \Delta\varphi$ can be analytically predicted as (see Fig. 4(b2), black curve):

$$
\delta\varphi' = \mathrm{sgn}(\beta_2)\left[\tan^{-1}\left(\frac{S_f(u_1) - S_f(u_2)}{C_f(u_1) - C_f(u_2)}\right) - \tan^{-1}\left(\frac{-\mathrm{sgn}(\beta_2) + S_f(u_1) - S_f(u_2)}{-\mathrm{sgn}(\beta_2) + C_f(u_1) - C_f(u_2)}\right)\right] \quad (19)
$$

The linear evolution for an initial phase offset of π and a duration $T_0$ of 40 ps is provided in Fig. 5(a1) and is compared with the temporal Arago spot (panel 5(a2)). We can note several features that are directly linked to the interference process that may exist between $\psi_A$ and $\psi_3$. First, due to constructive interference ($\delta\varphi$' being close to π), the temporal fluctuations that emerge on each side of the central area are significantly enhanced (see also Fig. 6(a) where an increase by 28 % of the peak power can be observed). Then the behavior in the central part is radically different, as displayed in panel 5(b1) where the longitudinal evolution of the intensity at $t = 0$ is summarized. Indeed, putting $Z = \pi T_0^2 / (z |\beta_2|)$, the central intensity is given by :

$$|a_\pi(0,Z)|^2 = 8\left[\left(C_f(Z) - \frac{1}{4}\right)^2 + \left(S_f(Z) - \frac{1}{4}\right)^2\right], \quad (20)$$

which should be compared with the central intensities of the considered aperture and the 1D Arago spot (purple and green lines respectively) that are given by [12] :

$$|\psi_3(0,Z)|^2 = 2\left[C_f^2(Z) + S_f^2(Z)\right]. \quad (21)$$

$$|\psi_A(0,Z)|^2 = 2\left[\left(C_f(Z) - \frac{1}{2}\right)^2 + \left(S_f(Z) - \frac{1}{2}\right)^2\right]. \quad (22)$$

All these quantities can be easily graphically predicted using the Cornu's spiral as directly linked to the distance between the point of curvilinear coordinate $Z$ and the fixed point (1/4, 1/4) for $|a_\pi|^2$, (0, 0) for $|\psi_3|^2$ and (1/2, 1/2) for $|\psi_A|^2$. Whereas the Arago spot has an amplitude that monotonously increases with propagation distance, the intensity observed at the center for the phase pattern experiences very strong fluctuations in the early stages of propagation that is a signature of the diffraction pattern of a rectangular pattern in the Fresnel regime (compare green and purple lines in panel b). The intensity at the center can be up to 2.8 times the average power of the illumining light, which is 58 % higher than the maximum of $|\psi_3(0)|^2$ taken alone. Propagation distance at which this maximum appears can be well predicted by a crossing point of two extrema located at $\pm T_0/2 \pm t_{max}$ which are given by Eq. (6). This observation provides scaling properties of the resulting pattern meaning that we can tailor the position of the maximum by carefully choosing the temporal extend of the initial phase jump. Note that the constructive interference process between $\psi_3$ and $\psi_A$ is not optimally efficient due to the very different intensity levels of the two waves. Moreover, the phase difference $\delta\varphi(0)$ that can be derived analytically as :

$$\delta\varphi'(t=0) = \tan^{-1}\left(\frac{1/2\,C_f(Z) - 1/2\,S_f(Z)}{C_f^2(Z) - 1/2\,\text{sgn}(\beta_2)C_f(Z) + S_f^2(Z) - 1/2\,\text{sgn}(\beta_2)S_f(Z)}\right) \quad (23)$$

is not an odd multiple of $\pi$ (see Fig. 5(b2) as well as Fig. 4(b2)). For asymptotic propagation distance and similarly to the Arago spot, the central intensity tends to 1. Regarding the first ripples that surround the central peaks (see Fig. 6(a), red line), we note that their intensity is enhanced compared to $|\psi_3|^2 + |\psi_A|^2$ (black line). Indeed, for the position of the lateral maxima at $t_{max} + T_0/2$, the interference between $\psi_3$ and $\psi_A$ is constructive, $\delta'\varphi$ being an odd multiple of $\pi$ after 10 km of linear propagation (see Fig. 4(b2)).

The influence of $\Delta\varphi$ is illustrated in Fig. 6 for a dispersive propagation distance of 10 km and we compare the pattern achieved for $\Delta\varphi = \pi$ (red line) and $\Delta\varphi = 1.37\pi$ (blue line). The initial phase offset significantly impacts the visibility of the fringes as well as the amplitude of the maximum of the oscillations. A more systematic study of the influence of $\Delta\varphi$ is provided in panel (b). The temporal intensity pattern is fully symmetric, whatever $\Delta\varphi$ is. We can make out that, after 10 km of propagation, the most pronounced dips surrounding the central peak are not achieved for $\Delta\varphi = \pi$ but for $\Delta\varphi = 1.37\ \pi$. The intensity of the central part is also strongly influenced by $\Delta\varphi$ as can also be seen in panel (b2) where we can make out that the peak intensity at $t = 0$ follows a sinusoidal evolution typical of a two wave interference process. Once again, $\delta\varphi'$ helps us to understand why $\Delta\varphi = \pi$ is not the optimum value to achieve the highest central peak-power. The optimum interference is obtained when $\Delta_T\varphi(0) = 2\pi\ n$ (with $n = 0, \pm 1, \pm 2 \ldots$), i.e. when $\Delta\varphi = 2\pi - \delta\varphi'(0) = 4.3$ rad. On the contrary, destructive interference is obtained for $\Delta\varphi = -\delta\varphi'(0) + 2\pi(2n-1)$, explaining the dip that tends to appear for $0 < \Delta\varphi < 0.4\pi$.

*Nonlinear propagation*

We next investigate the impact of nonlinearity on the pulse profile in presence of anomalous dispersion. Figure 7(a1) illustrates the longitudinal evolution of the intensity profile for an initial central shift of $\pi$ and an average power of 290 mW. The ripples existing for linear propagation have turned into well-isolated coherent structure reaching a peak power that can be as high as 6.1 times the average power after a propagation distance of 5.25 km. The amplitude of the peaks neighboring the central part is also enhanced by the focusing nonlinearity in the anomalous regime of dispersion. The details of the longitudinal evolution of the power at $t = 0$ are provided in panel (b1) of Fig. 5 (red solid line). Contrary to the evolution experimentally recorded in [24] where combination of higher power with anomalous dispersion lowered the intensity of the temporal Arago spot, the peak-power of the central peak is significantly increased after 3 km of propagation and can be here nearly doubled in presence of focusing nonlinearity. Details of the temporal intensity profile obtained at the point of maximum focusing, i.e. after 5.25 km, is provided in panel (b1) of Fig.7 where we can make out that the central part of the pulse is well described by the shape of a Peregrine breather [52] :

$$\psi_{PS}(t) = \frac{\sqrt{P_{PS}}}{3}\left(1 - \frac{4}{1 + 4(t/T_{PS})^2}\right), \tag{24}$$

with $T_{PS}$ the temporal duration of the Peregrine soliton and $P_{PS}$ its peak power linked by $T_{PS} = (\beta_2 / P_{PS}\, \gamma)^{1/2}$. Details of the phase profile at the point of maximum focusing (Fig. 7(c1)) also confirms the presence of typical phase shift of $\pi$ in the pedestals of the structure [53]. This once again stresses the intimate connection that exists between solitons over finite background structure of complex nonlinear dynamics and compression processes [54]. Note that the first lateral breathing structures that also experience growth and decay features can similarly have their central part adjusted at their point of maximum focusing (after 8.6 km of propagation) by the shape of a Peregrine breather.

The nonlinear evolution in the normal regime of propagation (Fig. 7(a2)) is very different. Indeed, contrary to the emergence of a strong localized structure, we observe the generation of a pair of identical black solitons. The intensity evolves towards a constant level (see also panel (b) of Fig. 5, blue solid line) tending to the power of the initial continuous wave. In the case of attenuation-free propagation, the phase and intensity profiles asymptotically reshape towards an ideal black soliton in perfect agreement with Eq. (15) as can be seen from Fig. 7(b2) for propagation distance of 10, 20 and 40 km. The phase profile is also marked by a typical phase shift of $\pi$ at the minimum of the dark soliton. One can note that the centers of the black solitons are not exactly located at $t = \pm \Delta T/2$ indicating that the structures have in their initial stage of reshaping a slight velocity. Once formed, in absence of perturbation and attenuation, the pair of dark solitons is stable (see the two dark parallel stripes in Fig. 7(b2)).

The influence of the initial average power on the resulting pattern is summarized in Fig. 8(a) for anomalous and normal dispersion. Compared to the results discussed in section 2.2 (see Fig. 3), we can note several differences regarding the evolution in the anomalous dispersion regime. First of all, we note that, contrary to the peak intensity of the lateral peaks that remains more or less constant with the input average power in the case of a single phase jump, we observe here some significant fluctuations in the peak power, denoting a breathing behavior. The temporal position of these structures is affected by the power, higher the power is, closer the structures are from the initial abrupt phase jumps. Regarding the pattern in the central part, contrary to [24], the evolution is not monotonic with power. We observe that a well-defined double peak structure appears for a well-chosen power (around 0.9 W), which is reminiscent from the recent investigation of the generation of a pair of pulses from an initial super-Gaussian pulse [55]. We can expect the initial temporal duration $T_0$ to be an efficient mean to control the number of structures in this central part [56] and that more complex interactions could be observed for longer durations [57]. On the contrary, in the normal regime of propagation, two black solitons are visible with temporal location in the vicinity of the initial phase offsets and a normalized intensity equals to 1 between these two phase shifts.

The pattern achieved for $\Delta \varphi = \pi/2$ is also provided in panel (b). Quite remarkably, the differences that were very pronounced in the case of a single phase jump are here attenuated in the anomalous regime of dispersion so that we retrieve, at least qualitatively, the different features we previously discussed for panel (a). However, in the normal regime of dispersion, we can note, that the dark solitons that emerge are not black and move one away from each other with an intensity pattern that is symmetric.

*Influence of the finite modulation bandwidth*

In order to end this discussion and to get perspectives on potential experimental demonstration, we take into account the finite bandwidth of the initial phase modulation. Indeed, whereas in spatial optics, use of very straight edges is technically feasible, temporal optics faces the practical limitations of the optoelectronics that will ultimately limit the steepness of the fronts applied on the initial continuous wave. In the article investigating the temporal Arago spot, we have shown using super-Gaussian intensity profiles instead of ideal rectangular one has only limited impact on the resulting temporal pattern. As a first approximation, we consider here that the bandwidth limitations affecting the generation of the phase profile can be modelled by Gaussian filter with a full width at half maximum of 80 GHz. The smoothened phase profile is shown in Fig. 9(a) for a modulation of $\pi$ and $-\pi$ (red solid and dashed lines respectively). The pattern observed after a propagation distance of 10 km is plotted in Fig. 9(b) where we note that major differences affect the linear propagation. Indeed, given the strong interference process that occurs near the phase jumps, any change in the phase profile severely impacts the emergence of the features we have previously identified. We also note that the pattern induced by a softened phase shift of $\pi$ strongly differs from the one emerging from a softened phase shift of $-\pi$, which can be explained by the impact of the temporal gradient present in the transition region. Once again, the exact identification of the nature of the coherent structures is challenging and requires dedicated analytical tools.

The changes are also significant when considering the propagation in presence of nonlinearity, as stressed by panel (c1) of Fig. 9 that should be compared to Fig. 8(a). In the anomalous regime of propagation, we do not observe the central doublet that was generated in the ideal case for a power around 1W. The differences are even more striking when considering the nonlinear propagation in the normal regime of dispersion. Whereas the ideal π-phase step leads to the generation of two black solitons, taking the bandwidth limitations, the dark solitons are now grey. They have a non-null velocity and move away from each other. Note that such a pattern may, to some extent, qualitatively recall the trends observed in linearly frequency modulated signal as analyzed in nonlinear optics [36, 58] based on Witham approaches initially used in fluids [59].

The nonlinear dynamics gets very different when considering a phase step of –π. Given the initial temporal chirp, the two grey solitons tend to move towards each other's and collide after 10 km for an initial power of 0.38W. Taking advantage of the finite bandwidth could therefore be a simple but efficient mean to generate experimentally a doublet of dark solitons with opposite velocities that could then collide [60]. The pattern observed in the central region for anomalous propagation is also drastically impacted by the sign of the phase offset. The initial perturbation turns into an expanding nonlinear oscillatory structure with a higher number of coherent structures being present in the central part and experiencing growth and decay cycles.

**Conclusion**

To conclude, we have studied the temporal intensity pattern emerging from the linear or nonlinear evolution of a single or double phase jump. We have provided an interpretation of the pattern in terms of interferences of the well-known diffractive patterns of a straight edge, strip and slit. A complete analytical framework has been provided in terms of Fresnel integrals for the case of purely dispersive evolution. This has enabled us to stress the similarity as well as the differences that exist in the pattern resulting from a phase shift and from an intensity modulation.

We have extended our study of the propagation to the case where Kerr nonlinearity impairs the nonlinear pattern. Various coherent nonlinear structures emerge from the pattern according to the regime of dispersion. Whereas dark solitons appear in normally dispersive fibers, the evolution in the anomalous regime of dispersion involves ultrashort breathing structures that may require the use of more advanced tools to be fully identified. Our results stress that phase jump can be extremely efficient to seed modulation-instability driven processes and should therefore stimulate new theoretical and experimental developments in the field of optical rogue waves. We have shown that, in this context, the analytical understanding of the dispersive evolution could provide some interesting clues and can benefit from the integrability properties of the focusing nonlinear Schrodinger equation.

The present work can be extended in many aspects. First, the initial wave is not restricted to a continuous fully coherent wave. Similarly to the spatial case, we can indeed consider the impact of a temporal phase shift imprinted on an initial pulse [23, 61, 62], on a wave that has an initial chirp [63, 64] or a wave that is only partially coherent [3]. With the progress of coherent transmissions, it is also possible to combine intensity and phase modulation, therefore mimicking a partially transparent phase object [22, 48, 65]. Moreover, it is possible to benefit from the vectorial properties of light in fibers to explore new degrees of freedom and other nonlinear structures [66]. We also believe that the present discussion can be of help to better understand the evolution of the vectorial shock waves we recently described [67] and to catch the way cross-phase modulation may affect through dispersion the temporal pattern of a continuous wave [68].

As the NLSE is a universal mathematical model that also accurate to describe wave propagation in other fields of physics such as Bose-Einstein condensates [69] or hydrodynamics [70], our conclusions dealing with the impact of phase jumps can be extended to the nonlinear evolution of water waves. Indeed, in recent years, the link between temporal optics and hydrodynamics has been a fruitful driving force in stimulating the understanding of coherent structures in focusing and defocusing regimes of nonlinear propagation [45, 53, 71, 72].

Acknowledgements : C F acknowledges the support by the Région Bourgogne-Franche-Comté and the Institut Universitaire de France. We sincerely thank Hervé Rigneault for stimulating initial discussions on the space/time analogy and Bertrand Kibler for insights about nonlinear coherent structures.

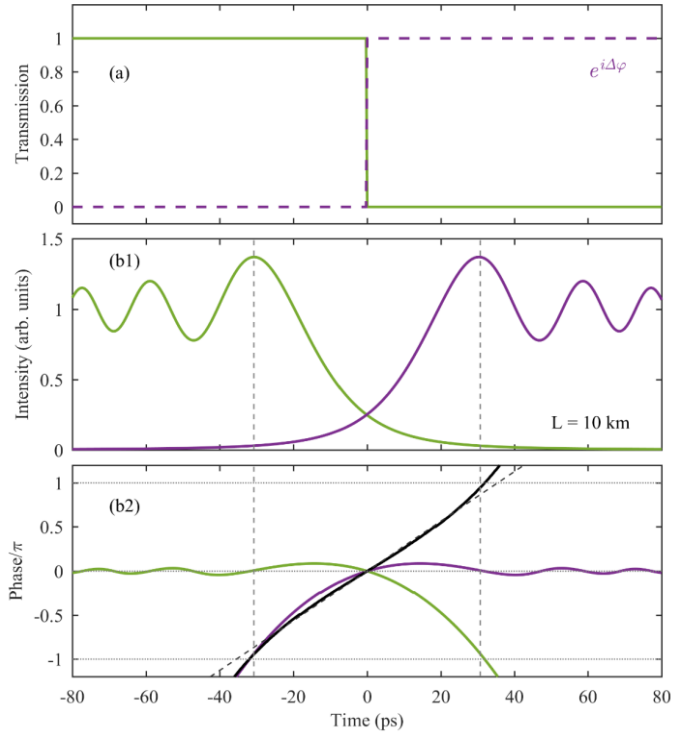

Figure 1. (a) Heaviside functions representing a 1D straight edge on the left-hand and the righthand sides with an imprinted phase offset of $\Delta\varphi$. (b1) The resulting intensity distribution of the diffracted light (b1) and the corresponding phase (b2) obtained after the propagation in 10 km optical fiber with $\beta_2 > 0$ in the linear regime. (properties of $\psi_1$ and $\psi_2$ are plotted with green and purple lines respectively). The exact phase difference $\delta\varphi$ (black solid line computed from Eq. (7)) is compared to a theoretical approximation given by Eq. (8) (black dashed line).

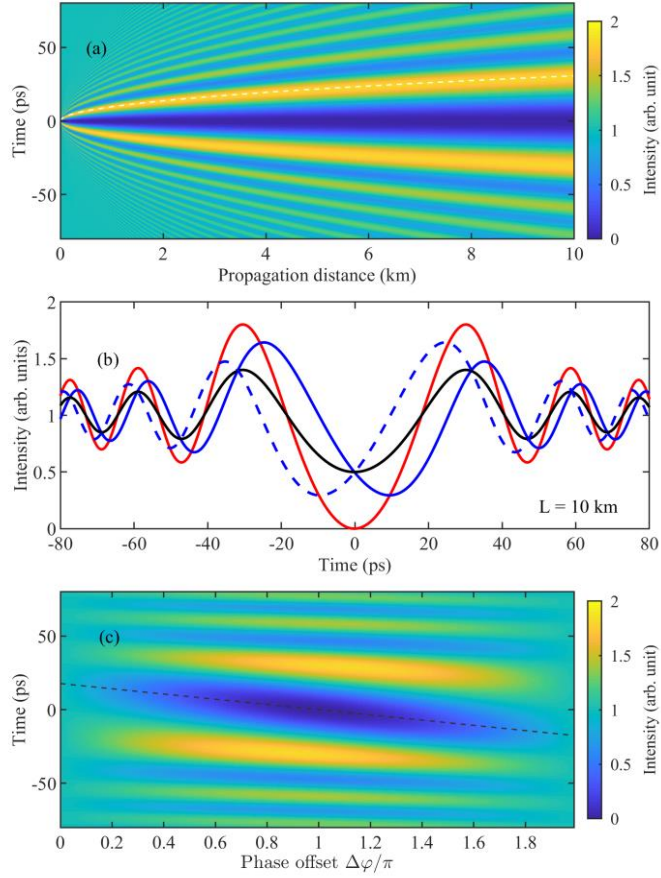

Figure 2. (a) Evolution of the temporal intensity profile in an optical fiber of light passed through a temporal single-step phase pattern with an offset of $\Delta\varphi = \pi$. The white dashed line marks the position of the first maxima given by Eq. (6). (b) The resulting intensity distribution at the output of 10 km fiber for $\Delta\varphi = \pi$ and $\Delta\varphi = \pi/2$ in the anomalous dispersion regime (red and blue solid lines, respectively) and for $\Delta\varphi = \pi/2$ in the normal dispersion regimes (blue dashed line). Black line depicts a sum of intensity profiles $|\psi_1|^2 + |\psi_2|^2$. (c) Intensity profiles at the end of anomalously dispersive fiber recorded at different values of the phase offset $\Delta\varphi$. Black dashed line marks position of the dip approximated by eq. (12). All intensity patterns computed numerically are in an perfect agreement with Eq. (9) and (10).

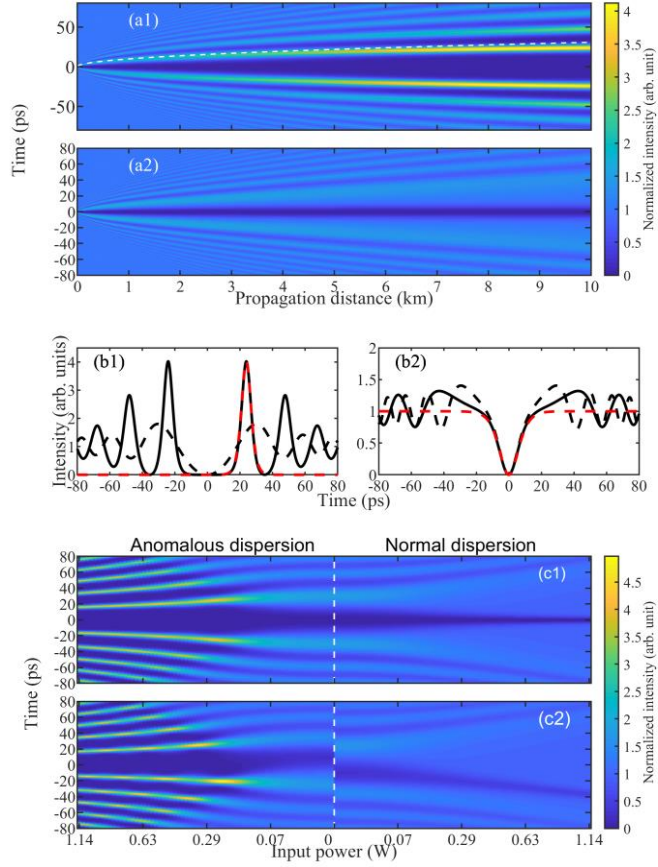

Figure 3. (a) Longitudinal nonlinear evolution of temporal intensity profile of the CW modulated by a single phase-jump profile with an offset of $\pi$ in anomalous (a1) and normal (a2) dispersion regimes ($\beta_2 = \pm 20$ ps$^2$ km$^{-1}$, $\gamma = 1.1$ W$^{-1}$ km$^{-1}$, $P_{av} = 0.29$ W). (b1) Details on the intensity profiles at the output of 10 km fiber with anomalous dispersion at the linear and nonlinear regimes (dashed and solid black lines respectively). Red line shows fit by a bright soliton given by Eq. (14). (b2) Close-up of the intensity profiles in normally dispersive fiber obtained after 6 and 10 km propagation distance (dashed and solid black lines respectively). Middle part of the structures is well described by the dark solution, Eq. (15) (red line). (c) Intensity distribution after a 10 km propagation with respect to the input power level for the phase offset $\Delta\varphi$ of $\pi$ and $\pi/2$ (panels 1 and 2 respectively).

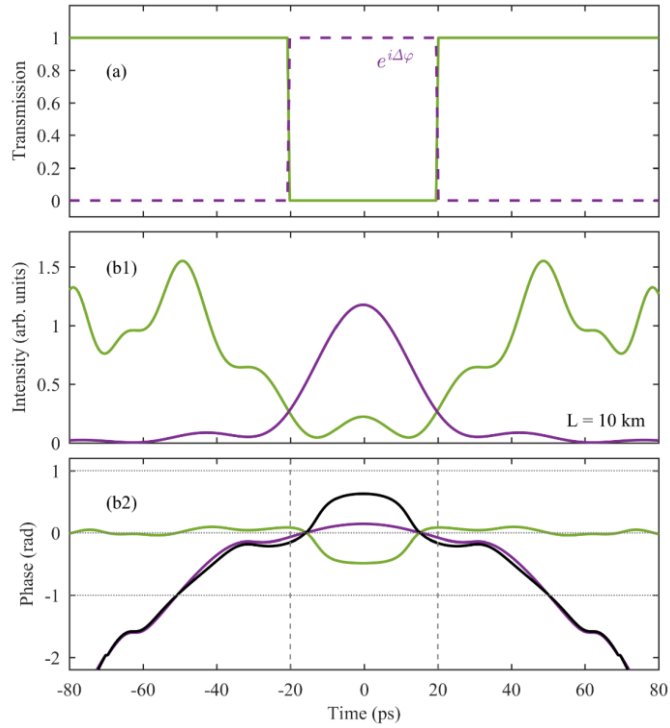

Figure 4. (a) Combination of two Heaviside functions as in Fig. 1(a) each shifted by ±20 ps ($\psi_A(t)$, green) and a rectangular profile of width of 40 ps ($\psi_3(t)$, purple) which has a phase offset of $\Delta\varphi$. (b) The corresponding intensity and phase profiles after a propagation in 10 km anomalously dispersive fiber (panels 1 and 2 respectively). Black line shows the a phase difference $\delta\varphi'$ obtained from Eq. (19).

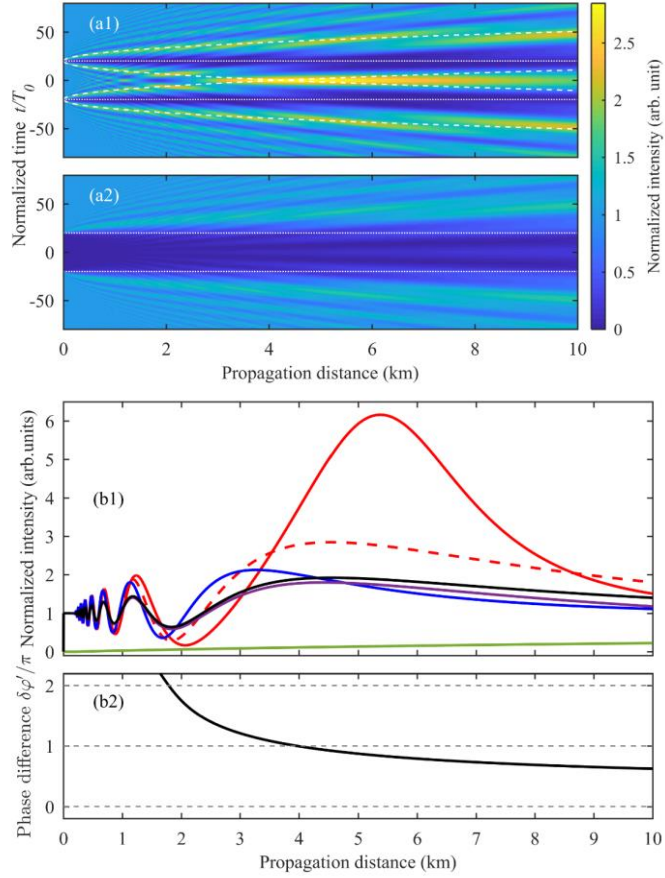

Figure 5. Longitudinal evolution of the temporal intensity profile of the CW modulated by the double phase-jump (a1) is compared to the temporal Arago spot (a2). Grey dotted lines mark positions of the imprinted patterns – phase or intensity jumps, respectively. White dashed lines represents position of the first maxima given by Eq. (6). (b1) Variation of the intensity and the phase $\delta\varphi'(0)$ (panels 1 and 2, respectively) at the center of the double phase-jump pattern. Intensity profiles are recorded for: $|\psi_A(t)|^2$, $|\psi_3(t)|^2$ and their sum (green, purple and black lines respectively) propagating in a linear regime with $\beta_2>0$; double phase-jump modulated wave propagating in fiber with $\beta_2>0$ at linear (red dashed line) and nonlinear (red solid line) regimes; the nonlinear evolution of the wave under the same conditions but with $\beta_2<0$ (blue solid line).

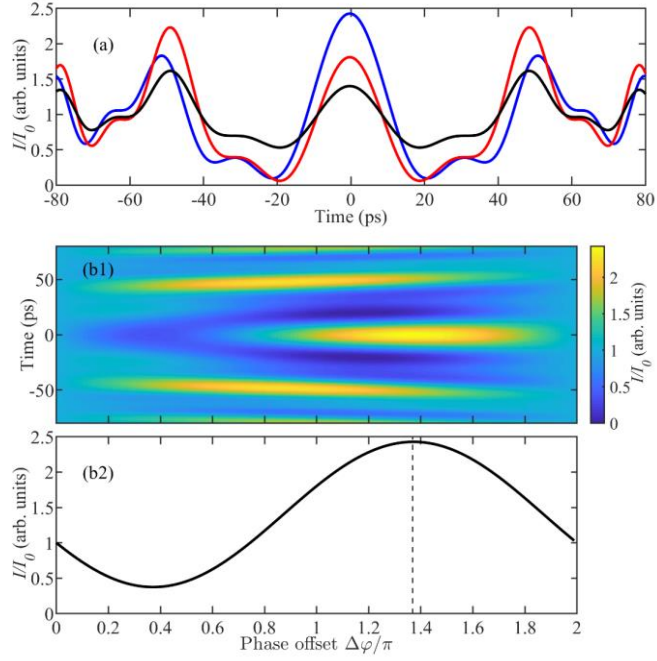

Figure 6. (a) Intensity distribution after the 10 km propagation in the anomalously dispersive fiber in the linear regime for the sum $|\psi_A(t)|^2 + |\psi_3(t)|^2$ (black), CW passed through the pattern with $\Delta\varphi = \pi$ (red) and $\Delta\varphi = 1.37\pi$ (blue). (b1) Intensity patterns simulated with the same fiber with varying phase offset. The presented waveforms are consistent with the theoretical predictions given by Eq. (17) and (18). (b2) Evolution of intensity in the central part is described by Eq. (20).

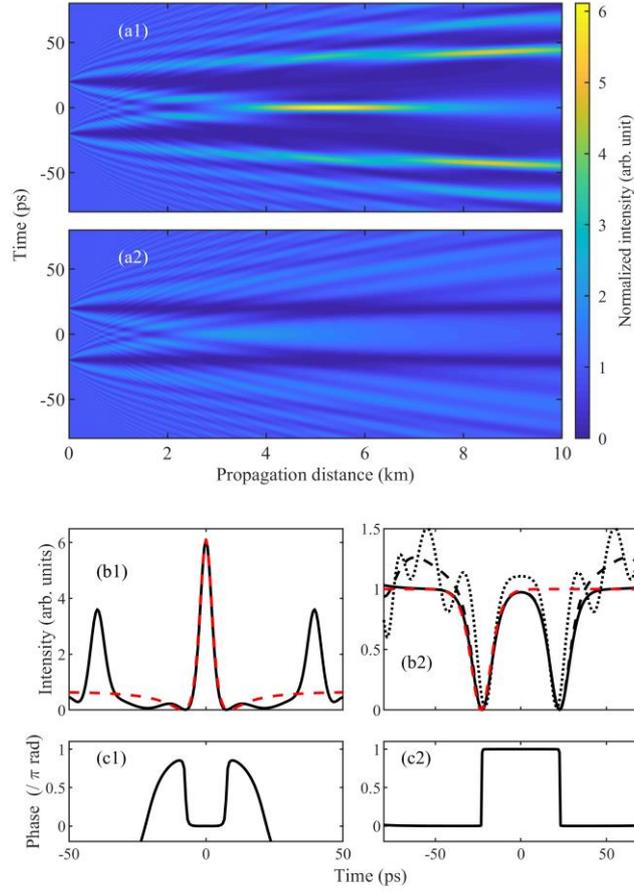

Figure 7. (a) Longitudinal nonlinear evolution of the intensity of the CW that is modulated by the double phase-jump with an offset of $\pi$ propagating in a fiber with anomalous and normal dispersion (panels 1 and 2 respectively) in the nonlinear regime. (b1) Waveform at a point of the maximum compression of panel (a1) (black line) is fitted by a Peregrine breather, Eq. (24) (red dashed line), whereas (c1) represents the corresponding phase profile of the resulting field. (b2) Intensity profiles obtained under the same conditions as in panel (a2) at different propagation lengths: after 10, 20 and 40 km of propagation (black dotted, dashed and solid lines respectively). Red dashed line shows a fit by the black soliton, Eq. (15).

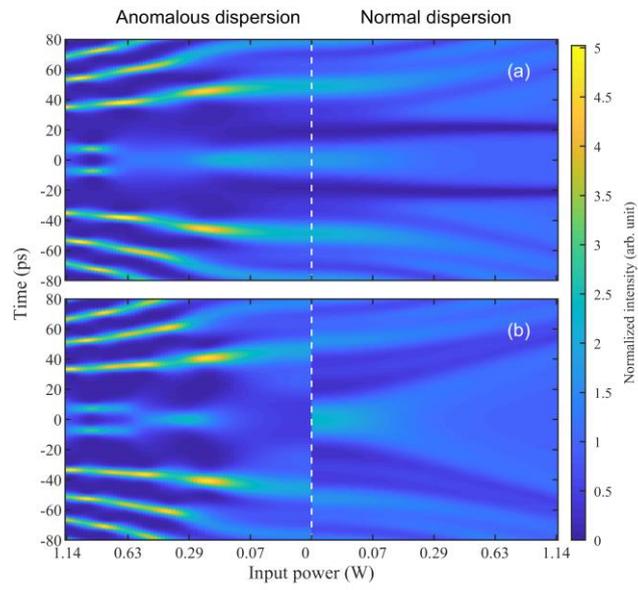

Figure 8. Temporal intensity profiles with respect to the initial power level of the CW modulated by the double phase-jump with an offset of $\pi$ and $\pi/2$ (panels (a) and (b) respectively) at the output of 10 km fiber.

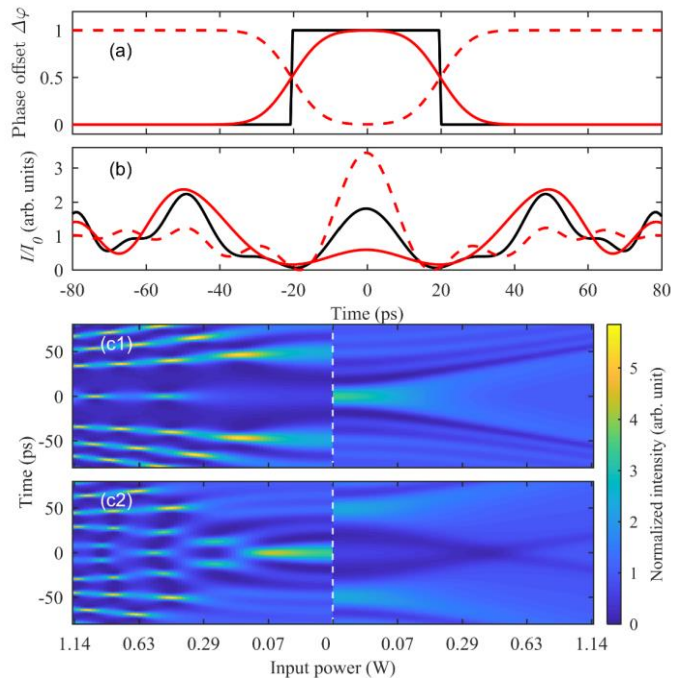

Figure 9. (a) Phase offset profiles: ideal double phase-jump (black line), shapes whose bandwidth is limited by a Gaussian profile with 80 GHz width with a depth of $\pi$ and $-\pi$ (red solid and dashed lines respectively). (b) The resulting intensity profiles recorded at the output of 10 km fiber with anomalous dispersion in the linear regime for the respected profiles in the panel (a). (c) Intensity mapped regarding the input power level for the phase profiles with a depth of $\pi$ and $-\pi$ (panels 1 and 2 respectively).